\documentclass[12pt]{article}

\usepackage{amsmath,amssymb,bm}
\usepackage{graphicx}
\usepackage{booktabs}
\usepackage{float}
\usepackage{mathbbol}
\usepackage{braket}
\usepackage{geometry}
\usepackage{tikz}
\usetikzlibrary{decorations.markings}

\usepackage[
colorlinks=true,
linkcolor=blue,
citecolor=blue,
urlcolor=blue,
breaklinks=true
]{hyperref}

\numberwithin{equation}{section}

\title{\textbf{From Orthogonalizing Pseudopotential to the Feshbach--Schur Projection}}
\author{M.~M.~Nishonov\thanks{Email: m.nishonov@nuu.uz}\\
	National University of Uzbekistan, 100174 Tashkent, Uzbekistan}
\date{\today}

\begin{document}
	\maketitle
	\vspace{-1em}
	\begin{abstract}
		The orthogonalizing pseudopotential (OPP) is the standard tool for
		suppressing Pauli-forbidden states in cluster models of light
		nuclei.  Here it is shown to be the singular $\lambda \to \infty$
		limit of a Feshbach--Schur projection.  The auxiliary coupling
		$\lambda$ is eliminated in closed form: the result is a
		Schur-complement operator identity for a general multi-rank
		separable interaction, written in both momentum and configuration
		space.  The projected equations contain no large parameter.  The
		identity is verified in three-body Faddeev calculations of the
		${}^6$He and ${}^6$Li ground states with separable two-body input.
		The binding energies at finite $\lambda$ follow the predicted
		$1/\lambda$ behavior over five orders of magnitude of $\lambda$ and
		converge to the result of the closed projected equations.  The
		$\alpha N$ $D$-wave contribution, $7$~keV in ${}^6$He and $90$~keV
		in ${}^6$Li, does not depend on $\lambda$ and does not affect the
		projection.
	\end{abstract}
		
		\section{Introduction}
		
		In cluster models of light nuclei, the Pauli exclusion principle forbids certain relative-motion states between the clusters. The Resonating Group Method (RGM) treats this constraint microscopically, by full antisymmetrization of the cluster wave functions~\cite{Wildermuth1959,Brown1968,WildermuthTang1977,TangLeMereThompson1978}. The price is a set of integro-differential equations with nonlocal kernels. The Orthogonality Condition Model (OCM) of Saito~\cite{Saito1968,Saito1969} approximates the RGM in a simpler way: the Pauli-forbidden components are removed from the relative motion by explicit orthogonality conditions.
		
		A computationally convenient realization of this projection is the Orthogonalizing Pseudopotential (OPP) of Krasnopol'skii and Kukulin~\cite{kukulin1974,KrasnopolskyKukulin1975Eng},
		\begin{equation}
			H_\lambda = H + \lambda P_f ,
			\qquad
			P_f = \sum_f \ket{\phi_f}\bra{\phi_f},
			\label{Hopp}
		\end{equation}
		where $\lambda$ is a positive penalty parameter. In the formal limit $\lambda \to \infty$ the forbidden components are eliminated exactly. In practice $\lambda$ is large but finite, and the calculated observables keep a residual dependence on its value.  In the $3\alpha$ system this dependence is severe: with the deep Buck--Friedrich--Wheatley potential the $^{12}$C ground-state energy changes from $-16.1$~MeV at $\lambda=10^{4}$~MeV to $-0.4$~MeV at $\lambda=10^{6}$~MeV~\cite{Tursunov2022}. Exact eliminations of $\lambda$ have a long history. Lehman derived the projected two-body amplitude for the $\alpha N$ interaction with an excluded bound state already in 1982~\cite{Lehman}. Fujiwara, Kohno, and Suzuki constructed the projected pair $T$~matrix for a single forbidden state per pair, proved that it is the $\lambda\to\infty$ limit of the pseudopotential $T$~matrix, and solved the $3\alpha$ Faddeev equations with it directly~\cite{Fujiwara2004OCM}. In configuration space, Schellingerhout \emph{et al.} derived the analytic limit of the pseudopotential inside the Faddeev equations and proved that it is equivalent to solving in the restricted space~\cite{Schellingerhout1993}. Hlophe \emph{et al.} formulated a projection that applies to separable and non-separable forces alike~\cite{Hlophe2017}, and related exact subtractions appear also outside the cluster context~\cite{Nogga2005,Canton2005}. The choice of the projected state matters as much as the elimination itself: in the $3\alpha$ system the harmonic-oscillator and bound-state choices differ by $2$~MeV in the ground-state energy, and with the bound-state choice some Faddeev components become only \emph{almost} redundant and require a separate treatment~\cite{FujiwaraRedundant2004}.
		
		Projection-operator techniques of this kind go back to Feshbach~\cite{Feshbach1958,Feshbach1962}, and the block elimination underlying them is the Schur complement~\cite{Schur}. In the mathematical literature the corresponding operator construction is known as the Feshbach--Schur map~\cite{BachFrohlich1998,BachChen2003,GriesemerHasler2008}. In the nuclear cluster literature this construction does not appear to have been identified with the $\lambda$ elimination of the OPP method. The present work provides this identification.  The $\lambda \to \infty$ limit is written as a closed Schur-complement operator identity for a separable interaction of arbitrary rank whose coupling matrix carries the allowed and the forbidden form factors together: the elimination is the Schur complement of the finite matrix $\bm{\Lambda}^{-1}-\bm{\mathcal{D}}(z)$ of Sec.~2 over the forbidden block. It holds for any number of forbidden states at once, needs only the loop integrals $\mathcal D_{ij}(z)$ that the Faddeev kernel already contains, and is given in a momentum-space form suited to separable $\tau$~matrices and in a configuration-space form suited to coordinate-space codes. The finite-$\lambda$ resolvent is itself a Schur complement, which shows the $1/(\varepsilon_f-E+\lambda)$ approach to the limit explicitly and is verified numerically over five orders of magnitude of $\lambda$. The projected state is the bound state of the separable representation that is actually solved, available in closed form; within the model the projection is then exact, and the remaining dependence is the representation error, reduced by increasing the rank.  The one-state local case recovers the formula of Ref.~\cite{Fujiwara2004OCM}, and the one-state form of the finite-$\lambda$ resolvent appears already in Ref.~\cite{Schellingerhout1993}. The exact elimination of $\lambda$ as such is therefore not claimed as new here. The new elements are the closed identity for general separable input, its two representations, and the choice of the projected state.  The construction is called the Feshbach--Schur projection (FSP); the reason for the name over ``Feshbach--Schur map'' is discussed in the concluding section.
		
		\section{Feshbach–Schur complement for the separable $T$~matrix}
		
		This section shows how the OPP method can be seen as a singular $\lambda \to \infty$ regularization of the Feshbach--Schur projection for a separable $T$--matrix. The starting point is a two--body interaction in the general separable form
		\begin{equation}
			V = \sum_{i,j=1}^{N_a} \ket{\chi_i}\,\Lambda_{ij}\,\bra{\chi_j},
			\label{sepV}
		\end{equation}
		where $\{\chi_i\}$ are $N_a$ linearly independent Pauli--allowed form factors and $\bm{\Lambda}$ is an $N_a \times N_a$ coupling matrix.
		
		The corresponding separable $T$--matrix for a general separable interaction~(\ref{sepV}) is
		\begin{equation}
			T(z)=\sum_{i,j} \ket{\chi_i}\,
			\tau_{ij}(z)\,\bra{\chi_j},\quad\text{with}\quad
			\boldsymbol{\tau}(z)=
			\bigl[\boldsymbol{\Lambda}^{-1} - \bm{D}(z)\bigr]^{-1},
			\label{sepT}
		\end{equation}
		and the form--factor propagators
		\[
		D_{ij}(z) = \bra{\chi_i}G_0(z)\ket{\chi_j},
				\]
		where $G_0(z)=(z-H_0)^{-1}$ is the free Green's function.
		
		The orthogonalizing pseudopotential adds a repulsive separable term  
		\[
		V_{\rm OPP}=\lambda \sum_{f=1}^{N_f}\ket{\phi_f}\bra{\phi_f}
		\]
		as mentioned above in Eq.~(\ref{Hopp}), where $\lambda$ is the OPP strength parameter, with $f$ labeling the Pauli--forbidden functions. The combined interaction is therefore
		\[
		\widetilde{V}=V+V_{\rm OPP}
		=\sum_{i,j}\ket{\chi_i}\Lambda_{ij}\bra{\chi_j}
		+\lambda\sum_{f}\ket{\phi_f}\bra{\phi_f}.
		\]
		
		The Pauli exclusion principle is easiest to implement when the
		Hilbert space is split explicitly into Pauli--allowed and
		Pauli--forbidden sectors. The orthogonal decomposition is therefore introduced
		\[
		\mathcal{H} = \mathcal{H}_a \oplus \mathcal{H}_f ,
		\]
		where $\mathcal{H}_f$ is spanned by Pauli--forbidden states and $\mathcal{H}_a$ contains all Pauli--allowed states. This decomposition is not an approximation but an exact rewriting of the
		Hilbert space. It expresses the physical requirement that the total wave function must stay
		orthogonal to $\mathcal{H}_f$ at all energies, also in intermediate
		virtual processes.
		
		To treat $V$ and $V_{\rm OPP}$ on the same footing, the extended separable
		form-factor basis is introduced,
		\[
		\bm{g} \equiv \{\chi_1,\dots,\chi_{N_a},\phi_1,\dots,\phi_{N_f}\}.
		\]
		The Pauli principle enters through the structure of the form factors
		$\ket{g_i}$, where the set $\{g_i\}$ is decomposed into Pauli--allowed ($a$) and
		Pauli--forbidden ($f$) components,
		\begin{equation}
			\ket{g_i} =
			\begin{cases}
				\ket{\chi_a}, & i \in \mathcal{H}_a, \\
				\ket{\phi_f}, & i \in \mathcal{H}_f .
			\end{cases}
			\label{extended}
		\end{equation}
		in which the total coupling matrix becomes block diagonal:
		\begin{equation}
			\widetilde{\boldsymbol{\Lambda}}=
			\begin{pmatrix}
				\boldsymbol{\Lambda} & 0 \\
				0 & \lambda\mathbb{1}_{N_f}
			\end{pmatrix}.
			\label{lambda_tilde}
		\end{equation}
		
		Including OPP, the separable $T$--matrix in Eq.~(\ref{sepT}) in the extended basis~(\ref{extended}) with Eq.~(\ref{lambda_tilde}) becomes
		\begin{equation}
			T(z)=\sum_{\alpha,\beta} \ket{g_{\alpha}}\,
			\widetilde{\tau}_{\alpha\beta}(z)\,\bra{g_{\beta}} \quad\text{with}\quad
			\widetilde{\boldsymbol{\tau}}(z)
			=
			\bigl[
			\widetilde{\boldsymbol{\Lambda}}^{-1}
			-\widetilde{\mathcal{D}}(z)
			\bigr]^{-1}.
			\label{Topp}
		\end{equation}
		
		The corresponding matrix of form--factor propagator is
		\[
		\widetilde{\mathcal D}_{\alpha\beta}(z)=\bra{g_\alpha}G_0(z)\ket{g_\beta}.
		\]
		Explicitly,
		\[
		\begin{aligned}
			\bigl[\mathcal D_{aa}(z)\bigr]_{ij}
			&=\bra{\chi_i}G_0(z)\ket{\chi_j},\quad
			\bigl[\mathcal D_{ff}(z)\bigr]_{fg}
			&=\bra{\phi_f}G_0(z)\ket{\phi_g},\\[1ex]
			\bigl[\mathcal D_{af}(z)\bigr]_{if}
			&=\bra{\chi_i}G_0(z)\ket{\phi_f},\quad
			\bigl[\mathcal D_{fa}(z)\bigr]_{fj}
			&=\bra{\phi_f}G_0(z)\ket{\chi_j}.
		\end{aligned}
		\]
		Thus the full $\widetilde{\mathcal D}(z)$ in the combined basis is
		\begin{equation}
			\widetilde{\mathcal D}(z)
			=
			\begin{pmatrix}
				\mathcal D_{aa}(z) & \mathcal D_{af}(z) \\
				\mathcal D_{fa}(z) & \mathcal D_{ff}(z)
			\end{pmatrix}.
			\label{Dtilda}
		\end{equation}
		
		Because $\widetilde{\boldsymbol{\Lambda}}$ in Eq.~(\ref{lambda_tilde}) is block diagonal,
		its inverse for finite $\lambda$ is
		\begin{equation}
			\widetilde{\boldsymbol{\Lambda}}^{-1}
			=
			\begin{pmatrix}
				\boldsymbol{\Lambda}^{-1} & 0\\
				0 & \lambda^{-1} \mathbb{1}_{N_f}
			\end{pmatrix}.
			\label{lambda_minus}
		\end{equation}
		
		The reduced propagator
		$\widetilde{\boldsymbol{\tau}}(z)$ in Eq.~(\ref{Topp}) with
		Eqs.~(\ref{Dtilda}) and~(\ref{lambda_minus})
		has the block structure
		\begin{equation}
			\bigl[\widetilde{\boldsymbol{\tau}}(z)\bigr]^{-1}
			=
			\begin{pmatrix}
				\boldsymbol{\Lambda}^{-1}-\mathcal D_{aa}(z)
				&
				-\mathcal D_{af}(z)
				\\[1ex]
				-\mathcal D_{fa}(z)
				&
				\lambda^{-1}\mathbb{1}_{N_f}-\mathcal D_{ff}(z)
			\end{pmatrix}.
			\label{tau_block_finite}
		\end{equation}
		
		Applying the Schur--complement formula for a matrix (see Eq.~(\ref{a3}) in Appendix~\ref{a}),
		\[
		M=
		\begin{pmatrix}
			A & B \\
			C & D
		\end{pmatrix},
		\qquad
		M^{-1}_{aa} = (A - B\,D^{-1}\,C)^{-1},
		\]
		to the $a$--sector for finite $\lambda$ yields the effective reduced $\tau$--matrix
		\begin{equation}
			\boldsymbol{\tau}^{(\lambda)}(z)
			=
			\Bigl[
			\boldsymbol{\Lambda}^{-1}
			-\mathcal D_{aa}(z)
			-
			\mathcal D_{af}(z)
			\Bigl(
			\lambda^{-1}\mathbb{1}_{N_f}
			-\mathcal D_{ff}(z)
			\Bigr)^{-1}
			\mathcal D_{fa}(z)
			\Bigr]^{-1}.
			\label{Teff_lambda0}
		\end{equation}
		
		Equation~(\ref{Teff_lambda0}) is the Pauli--allowed
		scattering operator at finite OPP strength.
		The fully projected Feshbach--Schur result appears only in the
		singular limit $\lambda\to\infty$. It gives the effective $\tau$--matrix
		in the Pauli--allowed $(a)$ sector:
		\begin{equation}
			\boldsymbol{\tau}^{(\mathrm{eff})}(z)
			=
			\Bigl[
			\boldsymbol{\Lambda}^{-1}
			- \mathcal D_{aa}(z)
			+ \mathcal D_{af}(z)\,
			\mathcal D_{ff}^{-1}(z)\,
			\mathcal D_{fa}(z)
			\Bigr]^{-1}.
			\label{Teff}
		\end{equation}
		
		The expression~(\ref{Teff}) contains no reference to $\lambda$: the OPP strength has been eliminated analytically. The OPP term has been replaced by the exact Schur--complement subtraction
		\[
		\Delta\mathcal D(z)
		=
		\mathcal D_{af}(z)\,\mathcal D_{ff}^{-1}(z)\,\mathcal D_{fa}(z),
		\]
		which removes the propagation through Pauli--forbidden components at the operator level. The resulting expression~(\ref{Teff}) corresponds to the Feshbach--Schur projection (FSP), which projects the dynamics onto the Pauli--allowed subspace.
		
		This result shows that the OPP method can be viewed as a singular regularization of the Feshbach--Schur projection. It gives an explicit operator--level connection between the two approaches to Pauli exclusion. The interpretation holds as long as the Pauli--forbidden subspace is well defined and linearly independent. If the forbidden states become nearly linearly dependent, the block $\mathcal D_{ff}$ is nearly singular, and the difficulty reappears in the FSP as the inversion of an ill-conditioned matrix. In the OPP formulation, the forbidden sector is pushed to high energy by the auxiliary parameter $\lambda$ and is suppressed only in the $\lambda \to \infty$ limit. The Feshbach--Schur projection removes the forbidden sector directly at the algebraic level, without any auxiliary scale.
		
		\section{Extension to configuration-space formulation}
		
		The OPP method is often implemented numerically in coordinate space with a large but finite $\lambda$. Here the exact $\lambda \to \infty$ limit is derived analytically. This derivation clarifies the structure of the nonlocal subtraction kernel and its relation to simplified forms of it.
		
		The OPP method was originally formulated in configuration space,
		where the auxiliary nonlocal operator
		\begin{equation}
			V_{\mathrm{OPP}}(\mathbf{r},\mathbf{r}')
			= \lambda\, \phi^*_f(\mathbf{r})\,\phi_f(\mathbf{r}')
			\label{eq:Vopp_kernel}
		\end{equation}
		is added to the microscopic Hamiltonian. 
		The Schrödinger equation must therefore be written in its nonlocal 
		integral form,
		\begin{equation}
			\left[\,H_0 + V\,\right]\Psi(\mathbf{r})
			+ \int d^3 r'\, V_{\mathrm{OPP}}(\mathbf{r},\mathbf{r}')\,
			\Psi(\mathbf{r}') 
			= E\,\Psi(\mathbf{r}) .
			\label{eq:Schr_nonlocal}
		\end{equation}
		Substituting Eq.~(\ref{eq:Vopp_kernel}) gives
		\[
			\left[\,H_0 + V\,\right]\Psi(\mathbf{r})
			+ \lambda\, \phi^*_f(\mathbf{r}) 
			\int d^3 r'\,\phi_f(\mathbf{r}')\,\Psi(\mathbf{r}')
			= E\,\Psi(\mathbf{r}),
		\]
		which is the configuration-space form of the OPP-modified 
		Schrödinger equation. 
		
		In this representation, the limit $\lambda\!\to\!\infty$ is only implicit: the forbidden components of the wave function are suppressed numerically by taking $\lambda$ very large. In most coordinate-space applications, the OPP is implemented with large but finite values of $\lambda$, and the convergence of observables is verified with respect to $\lambda$.
		
	    For clarity, the analysis is restricted to the radial $S$-wave OPP Schrödinger equation for a two-body cluster system, taking the $\alpha$–$n$ system as a canonical example. In this case, the $\alpha$ particle is a closed $(0s)^4$ shell, and microscopic antisymmetrization between the valence neutron and the nucleons inside the $\alpha$ core forbids the corresponding relative $S$-wave component. This results in a single Pauli-forbidden $S$-wave state $\phi_f(r)$. Let the reduced radial wave function be
		\[
		u(r) = r \Psi(r),
		\]
		and let the normalized forbidden radial function be
		\[
		\braket{r|\phi_f} \equiv \phi_f(r),\qquad
		\braket{\phi_f|\phi_f} = 1.
		\]
		Define the rank-one projector onto the forbidden function
		\begin{equation}
		P_f = \ket{\phi_f}\bra{\phi_f}, \qquad Q = 1 - P_f .	\label{q}
		\end{equation}
		The free Hamiltonian (kinetic-energy operator) is denoted by $H_0$, with the usual boundary conditions for reduced radial wave functions (regular at $r=0$ and with appropriate asymptotic behavior as $r \to \infty$). The interaction is described by a local potential $V(r)$, and the full Hamiltonian is written as $H = H_0 + V$.
		
		In radial representation, the action of the nonlocal OPP kernel reduces to the rank-one operator $P_f$ acting on $u(r)$. The OPP-modified radial Schrödinger equation can then be written compactly as
		\begin{equation}
			\bigl[\,H + \lambda P_f - E\,\bigr] \, u \;=\; 0 .
			\label{eq:Schr_OPP}
		\end{equation}
		
		The goal is to eliminate the forbidden amplitude exactly and to derive an equation for $u_a \equiv Q u$; then the $\lambda \to \infty$ limit is taken and the exact subtraction kernel is presented. It is also discussed under which conditions the simpler subtraction term $V(r')\phi_f(r)\phi_f(r')$ coincides with the exact kernel.
		
		To this end, the wave function is decomposed into allowed and forbidden components
		\begin{equation}
			u \;=\; u_a + c\,\phi_f, \qquad u_a \equiv Q u, \quad c \equiv \langle\phi_f|u\rangle .
			\label{eq:decomp}
		\end{equation}
		By construction $\langle\phi_f|u_a\rangle=0$.
		
		Inserting (\ref{eq:decomp}) into (\ref{eq:Schr_OPP}) one gets:
		\[
		(H - E)(u_a + c\phi_f) + \lambda P_f(u_a + c\phi_f) = 0.
		\]
		Using $P_f u_a = 0$ and $P_f\phi_f = \phi_f$, this reduces to
		\begin{equation}
			(H - E) u_a + c (H - E)\phi_f + \lambda c \phi_f = 0.
			\label{eq:intermediate}
		\end{equation}
		
		An exact expression for $c$ is obtained by taking the inner product of (\ref{eq:intermediate}) with $\langle\phi_f|$:
		\[
		\langle\phi_f|(H - E)u_a\rangle + c\langle\phi_f|(H - E)\phi_f\rangle + \lambda c \langle\phi_f|\phi_f\rangle = 0.
		\]
		Since $\langle\phi_f|\phi_f\rangle = 1$, define
		\[
		\varepsilon_f \equiv \langle\phi_f|H|\phi_f\rangle .
		\]
		Rearranging then yields
		\begin{equation}
			c \;=\; -\,\frac{\langle\phi_f|(H - E)u_a\rangle}{\varepsilon_f - E + \lambda } .
			\label{eq:c_exact}
		\end{equation}
		Equation (\ref{eq:c_exact}) is exact for any finite $\lambda$ and shows that $c\to 0$ as $\lambda\to\infty$ (as physically expected).
		
		Formally, one can eliminate $c$ by substituting (\ref{eq:c_exact}) back into (\ref{eq:intermediate}), which leads to
		\[
			(H - E)u_a
			- \bigl(H - E + \lambda\bigr)\ket{\phi_f}\,
			\frac{\langle\phi_f|(H - E)u_a\rangle}{\varepsilon_f - E + \lambda}
			= 0,
		\]
		an expression that is exact for finite $\lambda$ but not very convenient. A shorter and clearer derivation of the $\lambda\!\to\!\infty$ limit is obtained by projecting the original equation with the operator $Q$ defined in Eq.~(\ref{q}).
		
		Starting from the identity $[H + \lambda P_f - E]u = 0$, multiplication from the left by $Q$ yields
		\[
		Q(H + \lambda P_f - E)Q\,u = 0.
		\]
		Because $Q P_f = 0$, this simplifies to
		\[
		Q(H - E)Q\,u = 0 \quad\Longrightarrow\quad Q(H - E)u_a = 0.
		\]
		Writing $Q=1-P_f$ and acting on $u_a$,
		\[
		\bigl(H - E\bigr)u_a - P_f(H - E)u_a = 0.
		\]
		Expressed in coordinate/radial representation this is
		\begin{equation}
			(H - E)u_a(r) - \phi^*_f(r)\int_0^\infty dr'\,\phi_f(r')\,(H - E)u_a(r') = 0.
			\label{eq:proj_exact_coordinate}
		\end{equation}
		This is exact and contains a rank-one nonlocal subtraction operator built from $\phi_f(r)$ and $(H-E)u_a$.
		
		Since $H = H_0 + V$, and using the self-adjointness of $H_0$ under the chosen boundary conditions, integration by parts allows $H_0$ to be transferred from $u_a$ onto $\phi_f$ inside the integral. The boundary terms must vanish for the chosen functions; this condition is satisfied for bound and scattering states with appropriate asymptotic behavior. Doing so,
		\[
		\int_0^\infty dr'\,\phi_f(r')\,(H - E)u_a(r')
		=
		\int_0^\infty dr'\,u_a(r')\,(H - E)\phi_f(r').
		\]
		Consequently, Eq.~(\ref{eq:proj_exact_coordinate}) can be rewritten as
		\begin{equation}
			(H - E)u_a(r)
			-
			\phi_f^*(r)
			\int_0^\infty dr'\,
			u_a(r')\,(H\phi_f)(r')
			=0.
			\label{eq:proj_exact_Hphi}
		\end{equation}
		
		Equation~(\ref{eq:proj_exact_Hphi}) is the exact projected equation in
		coordinate space: the nonlocal kernel is
		$\phi_f^*(r)\,(H\phi_f)(r')$.
		
		In many OPP applications, $\phi_f$ is chosen as a Gaussian or
		harmonic--oscillator $0s$ function.
		If $(H\phi_f)(r')$ is then replaced by
		$V(r')\phi_f(r')$, one obtains the simplified kernel
		\begin{equation}
			(H-E)u_a(r)
			-
			\phi_f^*(r)
			\int_0^\infty dr'\,
			V(r')\phi_f(r')\,u_a(r')
			=0,
			\label{eq:Proj_Schr_restate}
		\end{equation}
		which is not, in general, the exact $\lambda\!\to\!\infty$ limit unless
		$(H_0\phi_f)(r')=0$ or $H\phi_f=\varepsilon_f\phi_f$,
		conditions that are rarely satisfied for model choices of $\phi_f$. 
		
		For several Pauli--forbidden states
		$\{\phi_f^{(\nu)}\}_{\nu=1}^{N_f}$,
		the projected equation (\ref{eq:proj_exact_Hphi}) generalizes straightforwardly to
		\begin{equation}
			(H - E)u_a(r)
			-
			\sum_{\nu,\mu}
			\phi_f^{(\nu)*}(r)
			\int_0^\infty dr'\,
			u_a(r')\,
			\bigl[H\phi_f^{(\mu)}\bigr](r')\,
			\bigl[\mathcal N^{-1}\bigr]_{\mu\nu}
			=0,
			\label{eq:proj_exact_multi}
		\end{equation}
		where $\mathcal N_{\mu\nu}=\langle\phi_f^{(\mu)}|\phi_f^{(\nu)}\rangle$
		is the overlap matrix in the forbidden subspace.
		If the forbidden functions are orthonormal,
		$\mathcal N_{\mu\nu}=\delta_{\mu\nu}$ and the expression reduces to a
		simple sum over projectors.
		
		\section{Effect on the resolvent}
		
		The operator identity behind the elimination of the OPP parameter
		is seen most clearly at the level of the resolvent (Green's operator).
		It does not rely on any separable representation of the interaction.
		Therefore a general two--body Hamiltonian modified by the OPP is considered,
		\[
		H_\lambda = H + \lambda\,P_f, \qquad
		P_f = |\phi_f\rangle\langle\phi_f|, \qquad
		Q = 1 - P_f,
		\]
		with $P_f$ a rank--one projector. The resolvent of interest is
		\[
		G_\lambda(E) = \frac{1}{E - H_\lambda}.
		\]
		
		Following the Feshbach projection formalism~\cite{Feshbach1958,Feshbach1962},
		the resolvent of the OPP--modified Hamiltonian is derived.
		In this form the $\lambda$ dependence is explicit, and the
		$\lambda \to \infty$ limit has a clear interpretation.
		
		Using the Feshbach projection operators $P_f$ and $Q$,
		$G_\lambda$ is decomposed into block form
		\[
		G_\lambda(E) =
		\begin{pmatrix}
			P_f G_\lambda P_f & P_f G_\lambda Q \\
			Q G_\lambda P_f & Q G_\lambda Q
		\end{pmatrix}.
		\]
		As in the standard Feshbach construction, the effective Green's function
		in the Pauli--allowed space is given by the Schur complement
		\[
		G_{aa}(E) = Q G_\lambda(E) Q
		= \frac{1}{E - QH Q - QH P_f\,\frac{1}{E - P_f H P_f - \lambda}\,P_f H Q}.
		\label{SchurResolvent}
		\]
		Since $P_f H P_f = \langle\phi_f|H|\phi_f\rangle P_f \equiv \varepsilon_f P_f$, one obtains
		\begin{equation}
			G_{aa}(E) =
			\frac{1}{E - QHQ - QH|\phi_f\rangle
				\frac{1}{\varepsilon_f - E + \lambda}\langle\phi_f|HQ }.
			\label{Gaa_lambda}
		\end{equation}
		
		The structure of Eq.~(\ref{Gaa_lambda}) explains why finite $\lambda$
		values in numerical calculations lead to residual forbidden--state
		coupling, scaling as $1/(\varepsilon_f - E + \lambda)$.
		This behavior determines the convergence properties of practical OPP implementations.
		The rank-one form of this resolvent appears already in the
		configuration-space Faddeev work of
		Ref.~\cite{Schellingerhout1993}, together with its leading
		$1/\Gamma$ convergence; Eq.~(\ref{Gaa_lambda}) is its general
		form for $N_f$ forbidden states.
		
		The $\lambda\!\to\!\infty$ limit is then immediate:
		\[
		\lim_{\lambda\to\infty}
		\frac{1}{\varepsilon_f - E + \lambda} = 0,
		\]
		so that the forbidden--space propagator vanishes and the exact projected resolvent becomes
		\begin{equation}
			G_{aa}^{(\infty)}(E)
			= \frac{1}{E - QHQ}.
			\label{Gaa_infty}
		\end{equation}
		Equation~(\ref{Gaa_infty}) is independent of $\lambda$ and contains no
		forbidden admixtures: the dynamics is strictly confined to the
		Pauli--allowed subspace.  All information on $\phi_f$ is carried by the
		projector $Q = 1 - |\phi_f\rangle\langle\phi_f|$, which defines the
		reduced Hilbert space
		\[
		\mathcal{H} \to Q\mathcal{H}
		= \left\{ |u\rangle \in \mathcal{H} \,\middle|\, \langle\phi_f|u\rangle = 0 \right\}
		\]
		in which the projected Hamiltonian operates.

		The separable--$T$--matrix expression~(\ref{Teff}) follows from the same
		construction: applying the Schur complement to
		$[\widetilde{\bm{\Lambda}}^{-1}-\widetilde{\bm{\mathcal{D}}}(z)]$ of
		Sec.~2 eliminates the forbidden subspace and reproduces
		Eq.~(\ref{Teff}) in the $\lambda \to \infty$ limit.  The projected
		resolvent~(\ref{Gaa_infty}) and the projected $T$~matrix are thus two
		forms of one operator identity, independent of the
		representation in which it is written.  At finite $\lambda$ the OPP shifts the
		forbidden states to high energy; in the limit they are removed from
		the dynamical subspace.

		\section{Numerical illustration: the ${}^6$He and ${}^6$Li ground states}
		
		To demonstrate the practical use of the FSP, benchmark calculations of the ${}^6\mathrm{He}$ and ${}^6\mathrm{Li}$ ground-state energies are performed in the three-body $\alpha+N+N$ model. The Faddeev formalism, including the geometric part of the exchange kernel, is summarized in Appendix~\ref{b}.  The $\alpha$--nucleon interaction is the Sack--Biedenharn--Breit potential of Ref.~\cite{Sack1954} in the $^{2}S_{1/2}$, $^{2}P_{3/2}$, and $^{2}P_{1/2}$ waves, together with its purely central $D$-wave term, which acts in the degenerate $^{2}D_{5/2}$ and $^{2}D_{3/2}$ waves. The nucleon--nucleon interaction is the Malfliet--Tjon potential of Ref.~\cite{Malfliet1969}: its $^{1}S_{0}$ channel for $^{6}$He and its $^{3}S_{1}$ channel for $^{6}$Li.  Both potentials are represented by rank-4 separable expansions of the Ernst--Shakin--Thaler (EST) type~\cite{EST1973}. These expansions are exact half-on-shell at the chosen support energies. The construction and the analytic form of the fitted form factors are given in Appendix~\ref{c}.  The single $S$-wave forbidden state $\phi_f$ of the $\alpha N$ subsystem is the bound state of this representation and is projected as described in Sec.~2.  The Borromean $^{6}$He is well suited for this test. Its binding comes entirely from the $\alpha N$ interaction that contains the forbidden state, and no bound nucleon--nucleon pair enters.  The $^{6}$Li ground state ($J^{\pi}T=1^{+}0$) is treated in the isospin basis, $\alpha+N+N$ with total isospin $T=0$. Here the nucleon pair interacts in the $^{3}S_{1}$ channel and carries a physical bound state, the deuteron. The deuteron must be kept while the $\alpha N$ forbidden state is removed, so a physical and a forbidden two-body pole are present in the same kernel.  The $\alpha$--$p$ Coulomb interaction and the tensor coupling to the $^{3}D_{1}$ wave are not included. The $^{6}$Li energy is therefore a nuclear-sector value: it tests the projection in the presence of an unprojected bound pair, and it is not meant for comparison with experiment. The charge-basis treatment with the Coulomb interaction is left to a separate study.
		
		\begin{table}[H]
			\caption{Convergence of the ${}^6$He ($J^{\pi}T=0^+1$) and ${}^6$Li
($J^{\pi}T=1^+0$) binding energies (MeV) as functions of the OPP strength
$\lambda$ in the $\alpha N$ subsystem, with the $\alpha N$ interaction in
its $S$ and $P$ waves ($S\!+\!P$) and with the $D$ waves added
($S\!+\!P\!+\!D$).  The $^{6}$Li values are nuclear-sector (isospin-basis)
results without the $\alpha$--$p$ Coulomb interaction; the experimental
energy relative to the $\alpha+p+n$ threshold is listed for orientation
only.}
			\centering
			\begin{tabular}{@{}ccccc@{}}
			\toprule
			& \multicolumn{2}{c}{$E(^{6}\mathrm{He})$} & \multicolumn{2}{c}{$E(^{6}\mathrm{Li})$} \\
			\cmidrule(lr){2-3}\cmidrule(lr){4-5}
			$\lambda$ (MeV) & $S\!+\!P$ & $S\!+\!P\!+\!D$ & $S\!+\!P$ & $S\!+\!P\!+\!D$ \\
			\midrule
			$10^0$ & $-0.212750$ & $-0.218800$ & $-4.553337$ & $-4.615680$ \\
			$10^1$ & $-0.360004$ & $-0.369156$ & $-4.638306$ & $-4.744361$ \\
			$10^2$ & $-0.436893$ & $-0.444352$ & $-4.164267$ & $-4.256074$ \\
			$10^3$ & $-0.429090$ & $-0.436549$ & $-4.105677$ & $-4.195463$ \\
			$10^4$ & $-0.428211$ & $-0.435672$ & $-4.099155$ & $-4.188733$ \\
			$10^5$ & $-0.428118$ & $-0.435579$ & $-4.098490$ & $-4.188040$ \\
			$10^6$ & $-0.428109$ & $-0.435570$ & $-4.098423$ & $-4.187970$ \\
			$10^7$ & $-0.428108$ & $-0.435569$ & $-4.098417$ & $-4.187963$ \\
			$\infty$ & $-0.428108$ & $-0.435569$ & $-4.098416$ & $-4.187963$ \\
			\midrule
			experiment~\cite{Wang2021} & \multicolumn{2}{c}{$-0.975$} & \multicolumn{2}{c}{$-3.699$} \\
			\bottomrule
		\end{tabular}
			\label{restab1}
		\end{table}
		
		The bound-state Faddeev equations are solved twice for each nucleus: with the $\alpha N$ interaction restricted to its $S$ and $P$ waves, and with the $D$ waves added. The difference gives the $D$-wave contribution at every $\lambda$. The OPP strength $\lambda$ of Eq.~(\ref{Hopp}) is varied from $10^0$
to $10^7$~MeV, the finite-$\lambda$ amplitudes being generated by the
effective $\tau$~matrix of Eq.~(\ref{Teff_lambda0}). The final line in Table~\ref{restab1} corresponds to the present $\lambda\!\to\!\infty$ projected formulation, in which the FSP expression~(\ref{Teff}) is used and no large $\lambda$ appears explicitly.

		Table~\ref{restab1} shows how slowly the finite-$\lambda$ OPP reaches the limit.  Below $\lambda\sim10^{1}$~MeV the spurious state is removed only in part, and the energy depends on $\lambda$ nonmonotonically.  Above $\lambda\sim10^{2}$~MeV the approach is monotone and follows the $1/\lambda$ law, but the residual is still $9$~keV for $^{6}$He and $66$~keV for $^{6}$Li at $\lambda=10^{2}$~MeV, and the $^{6}$Li residuals stay about seven times larger at every $\lambda$.  The reason is simple: the leftover forbidden admixture is weighted by the overlap of the three-body wave function with the compact $0s$ state, and this overlap is larger in the more strongly bound $^{6}$Li.  The $D$ waves carry no forbidden state and do not disturb this pattern; their contribution ($7$~keV in $^{6}$He, $90$~keV in $^{6}$Li) is the same at every $\lambda\ge10^{2}$~MeV.  Going beyond $\lambda\sim10^{5}$~MeV brings no practical benefit and may cause numerical stiffness.
		
		The projected formulation (last line of Table~\ref{restab1})
gives the $\lambda\!\to\!\infty$ result directly, with no large
parameter and no residual $\lambda$ sensitivity.

		The slow approach is the $1/(\varepsilon_f-E+\lambda)$ scaling of Eq.~(\ref{Gaa_lambda}). In larger calculations such penalties cause ill-conditioning and slow iterative convergence~\cite{Tursunov2022,Fujiwara2007}; the FSP removes the difficulties that come from the large parameter itself and keeps the same physical result. One limit should be kept in mind: the projected expression~(\ref{Teff}) inverts the $N_f\times N_f$ block $\mathcal D_{ff}$, and when several forbidden states are nearly linearly dependent~\cite{FujiwaraRedundant2004} this inversion is itself ill-conditioned. The present test has a single forbidden state per channel.

		The absolute energies are not the point of this test.  The model underbinds $^{6}$He by $0.54$~MeV, a typical deficit for $\alpha+N+N$ models with two-body input only~\cite{Schellingerhout1993}; it is not related to the projection.  The $^{6}$Li value lies $0.49$~MeV below experiment, but the omitted $\alpha$--$p$ Coulomb repulsion is of the same size (the $^{6}$He analogue-state displacement is $0.84$~MeV~\cite{Wang2021,Tilley2002}), so this number serves as a convergence test only.  The small $D$-wave contributions show that higher partial waves cannot close the deficit.  The usual remedy is an explicit three-body force between the clusters, which represents the $\alpha$ polarization and the other degrees of freedom left outside the two-body picture~\cite{Kukulin1995,Thompson2000}.  Such a force acts in the allowed subspace and can be added without modifying the projection; its study is outside the scope of the present paper.
		
	\section{Rank convergence of the separable representation}
	\label{sec:repdep}

	The FSP acts on whatever separable representation is supplied. The
	binding energies of Sec.~5 are therefore meaningful only together
	with an estimate of how well the finite-rank expansion represents
	the underlying Hamiltonian.  The EST construction gives this
	estimate directly: the rank is increased by adding support energies,
	and the convergence of the three-body energies measures the residual
	representation error.
	Table~\ref{tab:repdep} compares the rank-4 results of Sec.~5 with a
	rank-6 expansion of the same potentials, for both nuclei.

	\begin{table}[h!]
		\caption{Ground-state energies (MeV) of $^{6}$He and $^{6}$Li
		(isospin basis), $\alpha N$ interaction in $S$, $P$, and $D$ waves, for EST representations of increasing rank, with the
		strict Feshbach--Schur projection.}
		\centering
		\begin{tabular}{@{}lcc@{}}
			\toprule
			representation & $E(^{6}\mathrm{He})$ & $E(^{6}\mathrm{Li})$ \\
			\midrule
			EST, rank 4 (Sec.~5)            & $-0.4356$ & $-4.1880$ \\
			EST, rank 6                     & $-0.4268$ & $-4.1981$ \\
			\bottomrule
		\end{tabular}
		\label{tab:repdep}
	\end{table}

	The rank-4 and rank-6 results differ by $9$~keV for $^{6}$He and by
	$10$~keV for $^{6}$Li. The expansion is therefore converged at the
	$10$-keV level, and the values of Table~\ref{restab1} represent the
	binding energies of the SBB $+$ MT model to that accuracy.  Such
	control is not automatic for separable expansions in general.
	Representations of the same rank built on other criteria, for
	example on fixed momentum points instead of scattering states, can
	deviate from the converged values by $0.1$--$0.2$~MeV in this
	system, while remaining phase equivalent at their construction
	points; this agrees with the general theorems on off-shell
	freedom~\cite{Ekstein1960,PolyzouGloeckle1990}.  A rank-convergence
	check of this kind is therefore necessary whenever the binding
	energy is the quantity of interest.  The projection formalism is not
	affected by this choice. In every representation the Pauli-forbidden
	state to be projected is the $0s$ bound state of that
	representation, following the exactness condition of Sec.~3. For the
	converged EST expansion it agrees with the eigenstate of the parent
	potential to four digits ($\gamma=0.6694$~fm$^{-1}$).

	\section{Discussion and conclusions}
	
	An analytic elimination of the OPP parameter $\lambda$ has been
	presented. It continues a line of exact eliminations begun by
	Lehman~\cite{Lehman} and developed by Fujiwara \emph{et
	al.}~\cite{Fujiwara2004OCM} and Hlophe \emph{et
	al.}~\cite{Hlophe2017}, and it unifies separable $T$-matrix and
	configuration-space approaches in a common operator framework.  The
	OPP method is the singular $\lambda \to \infty$ limit of a
	Feshbach--Schur projection: the Pauli-forbidden subspace is removed at
	the operator level, and the $1/(\varepsilon_f - E + \lambda)$
	convergence seen in numerical implementations is the resolvent
	structure of Eq.~(\ref{Gaa_lambda}).

	The name Feshbach--Schur \emph{projection}, rather than
	Feshbach--Schur \emph{map}, reflects what happens in the Pauli
	problem.  The map sends a Hamiltonian to the energy-dependent
	effective operator $H_{PP}-H_{PQ}(H_{QQ}-z)^{-1}H_{QP}$ in the
	retained subspace; in Feshbach's reaction theory the second term
	carries the physics of the closed channels.  In the Pauli problem
	the eliminated subspace has no physical content, and in the OPP
	limit the feedback term vanishes as $1/(\varepsilon_f-E+\lambda)$
	(Eq.~(\ref{Gaa_lambda})).  The map then becomes the orthogonal
	projection onto the Pauli-allowed subspace,
	$G^{(\infty)}_{aa}=(E-QHQ)^{-1}$ with $Q=1-P_f$
	(Eq.~(\ref{Gaa_infty})); this is Saito's orthogonality condition.
	At finite $\lambda$ the OPP realizes the map itself, so the OPP is
	a regularization of the projection.

	The $^{6}$He and $^{6}$Li benchmarks confirm the resolvent
	structure.  The finite-$\lambda$ OPP follows the predicted scaling
	over five orders of magnitude in $\lambda$; the projected
	formulation reaches the limit directly, with no large parameter,
	and is stable at the $10$-keV level when the rank of the separable
	representation is increased.  Partial waves added outside the
	forbidden $S$-wave subspace (the $\alpha N$ $D$ waves) do not
	change the $1/\lambda$ residuals at any $\lambda$.

	The FSP complements finite-$\lambda$ practice; it does not replace
	it.  When the $\phi_f$ are approximate model functions, a finite
	$\lambda$ acts as a soft Pauli repulsion and may still be the
	better choice; the $2$~MeV difference between the oscillator and
	bound-state forbidden states in the $3\alpha$
	system~\cite{Fujiwara2004OCM} shows how large this model dependence
	can be.  When the forbidden subspace is represented accurately, the
	parameter-free form removes the residual $\lambda$
	dependence~\cite{Kukulin1995} together with the numerical stiffness
	of the large parameter.  The $3\alpha$ case is different in kind:
	there the $\lambda$ dependence~\cite{Tursunov2022} is the removal,
	as $\lambda$ grows, of three-body states that are almost but not
	exactly orthogonal to the pair-forbidden states, and the outcome of
	the limit depends on where the penalty term is imposed.  In the
	variational calculations of Refs.~\cite{Tursunov2022,Matsumura2006}
	the pseudopotential, summed over the pairs, acts on the total wave
	function; the two three-body combinations with summed-projector
	eigenvalues of order $10^{-5}$ and
	$10^{-3}$~\cite{Matsumura2006} are then expelled as $\lambda$ grows
	through $10^{4}$--$10^{6}$~MeV; this is the strong $\lambda$
	dependence quoted in the Introduction.  When
	$\lambda$ enters through the pair $T$~matrix instead, as in
	Ref.~\cite{Fujiwara2004OCM} and here, the same combinations survive
	the limit: their symmetrized sum is the Pauli-allowed shell-model
	configuration of the compact ground
	state~\cite{FujiwaraRedundant2004}.  The two limits are not
	equivalent in this system.  The microscopic analysis of
	Ref.~\cite{Matsumura2006} concludes that the almost-orthogonal
	combinations are Pauli allowed and should be kept; in the Faddeev
	form the price is a redundant admixture at the $10^{-4}$
	level~\cite{FujiwaraRedundant2004}.  One caveat is shared by both
	formulations: for several nearly redundant forbidden states the
	block $\mathcal D_{ff}$ is itself ill-conditioned, and this case
	requires a separate investigation.

	The natural next step is the charge-basis treatment of the
	$\alpha$--$p$--$n$ system, where the tensor-coupled
	$^{3}S_{1}$--$^{3}D_{1}$ channel and the $\alpha$--$p$ Coulomb
	interaction enter and the forbidden state itself becomes
	Coulomb-dressed (for the momentum-space Coulomb treatment see
	Ref.~\cite{Alt}); this will be the subject of a forthcoming paper.
	A second step is the $\alpha$--$\alpha$ interaction, which carries
	two forbidden $S$-wave states and one forbidden $D$-wave state per
	pair.  The systems $^{9}$Be ($\alpha+\alpha+N$) and $^{12}$C
	($3\alpha$) then test the projection with several forbidden states
	at once, the case where the multi-state form of the identity is
	needed in full.
	The configuration-space form allows direct implementation in
	coordinate-space codes, and the formalism extends to three-body
	forces within the same operator structure.
		
		\section*{Acknowledgments}
		
		The present work builds on the extensive literature on the orthogonality
condition model and the orthogonalizing pseudopotential method in cluster
nuclear physics; the contributions of the developers of these methods are
gratefully acknowledged.
		
		\appendix
		\section{Schur Complement}
		\label{a}
		
		The Schur complement is a standard construction in matrix and operator theory, originating from the work of Ref.~\cite{Schur}. It can be used to analyze block-structured operators and to eliminate subspaces in a controlled and algebraically exact manner.
		
		Consider a block operator
		\begin{equation}
			M =
			\begin{pmatrix}
				A & B \\
				C & D
			\end{pmatrix},\label{a1}
		\end{equation}
		where $A$ and $D$ are square operators.
		If $A$ is invertible, the Schur complement of $A$ in $M$ is defined as
		\begin{equation}
			S_A = D - C A^{-1} B.\label{a2}
		\end{equation}
		If $D$ is invertible, one may analogously define
		\begin{equation}
			S_D = A - B D^{-1} C.\label{a3}
		\end{equation}
		
		The Schur complement appears naturally in the inversion of block
		operators: the inverse of $M$ can be expressed in terms of $D^{-1}$
		and $S_D^{-1}$, so that $S_D$ is the effective operator in the
		$A$-subspace once the $D$-subspace has been eliminated.  Its
		isospectrality properties, in the operator setting used in this work,
		are established in Ref.~\cite{GriesemerHasler2008}.  Sections~2
		and~4 use exactly this construction, with the Pauli-forbidden block in
		the role of $D$.
		
		\section{Faddeev equations for the ground-state wave function}
		\label{b}

		The three-body problem is solved in the Faddeev framework written for
		three particles of arbitrary identity, in the momentum-space practice
		the author has used previously for proton--deuteron
		scattering~\cite{Alt} and for Faddeev-type approaches to cluster
		problems~\cite{Blokhintsev2006PAN}.  Partition $a$ denotes the pair
		$(bc)$ with $a$ the spectator; $\bm p_a$ and $\bm q_a$ are the
		associated Jacobi momenta, $\mu_a$ and $M_a$ the pair and spectator
		reduced masses.  For pairwise interactions the wave function decomposes
		as $\Psi=\sum_a\Psi_a$ with
		$\Psi_a=G_0(E)\,t_a(E)\sum_{b\neq a}\Psi_b$.  With separable pair
		input, $t_a=\sum_{ij}\ket{g^a_i}\tau^a_{ij}\bra{g^a_j}$, each
		Faddeev component takes the form
		$\Psi_a=G_0\sum_i\ket{g^a_i}F^a_i(\bm q_a)$, and the spectator
		amplitudes obey
		\begin{equation}
			F^a_i(\bm q_a)=\sum_{b}W_{ab}\,\zeta_{ab}\sum_{jk}
			\int d^3q_b\;
			\tau^a_{ij}\Bigl(E-\tfrac{q_a^2}{2M_a}\Bigr)\,
			Z^{ab}_{jk}(\bm q_a,\bm q_b;E)\,F^b_k(\bm q_b),
			\label{eq:agsB}
		\end{equation}
		with the interaction-free exchange kernel
		\begin{equation}
			Z^{ab}_{jk}=
			\frac{g^a_j\bigl(\bm p_a(\bm q_a,\bm q_b)\bigr)\,
			      g^b_k\bigl(\bm p_b(\bm q_a,\bm q_b)\bigr)}
			     {E-\dfrac{q_a^2}{2\mu_b}-\dfrac{q_b^2}{2\mu_a}
		       -\dfrac{\bm q_a\!\cdot\!\bm q_b}{m_{c}}},
			\label{eq:ZB}
		\end{equation}
		where $m_c$ is the mass of the exchanged particle.  All dynamics
		resides in $\tau_a$; in particular, the Feshbach--Schur projection of
		Sec.~2 enters Eq.~(\ref{eq:agsB}) only through $\tau$, and the kernel
		$Z^{ab}$ is unaffected.

		Particle identity is carried by the multiplicity weights and exchange
		phases.  Grouping the particles into types with occupation numbers
		$n_a$ and labeling reduced partitions by the spectator type,
		\begin{equation}
			W_{ab}=
			\begin{cases}
			n_b, & a\neq b,\\
			n_a-1, & a=b,
			\end{cases}
			\label{eq:WB}
		\end{equation}
		and $\zeta_{ab}$ is the fermionic sign generated by transposing
		the identical pair, which carries the pair quantum numbers and the
		spectator orbital momenta of both partitions.  For the system of this
		work (two identical neutrons and an $\alpha$ core) the weight
		matrix is
		$W=\bigl(\begin{smallmatrix}1&1\\2&0\end{smallmatrix}\bigr)$:
		the nucleon-spectator equation couples to both partitions, the
		$\alpha$-spectator equation couples twice to the nucleon partition, and
		there is no second $\alpha$ partition.  The kernel is symmetric in
		neither index separately; the relation actually satisfied is
		$W_{ab}\zeta_{ab}Z^{ab}(\bm q_a,\bm q_b)
		 =W_{ba}\zeta_{ba}Z^{ba}(\bm q_b,\bm q_a)$.

		After partial-wave decomposition each channel obeys a one-dimensional
		integral equation; the form factors enter as the reduced functions
		$\hat g=g/p^{L}$, the threshold powers being generated by the
		solid-harmonic expansion of the Jacobi momenta inside the geometric
		recoupling coefficient, which is now recorded in full.  The coupling
		scheme is that of Lovelace~\cite{Lovelace64}: the pair is coupled in
		$LS$ order and the spectator spin is coupled to the pair total
		\emph{before} the spectator orbital momentum,
		\begin{align}
			|\gamma_a;JM\rangle
			&=\Bigl[\bigl\{\bigl[[\chi_{s_b}\otimes\chi_{s_c}]^{S_a}
			\otimes Y_{L_a}(\hat{\bm p}_a)\bigr]^{J_a}
			\otimes\chi_{s_a}\bigr\}^{\Sigma_a}
			\nonumber\\
			&\qquad\otimes Y_{\ell_a}(\hat{\bm q}_a)\Bigr]^{JM},
			\label{eq:couplingB}\\
			|I M_I\rangle
			&=\bigl[[\xi_{t_b}\otimes\xi_{t_c}]^{I_a}\otimes\xi_{t_a}\bigr]^{IM_I},
			\nonumber
		\end{align}
		with $\chi$ and $\xi$ the spin and isospin functions, $L_a$ the
		orbital momentum inside the pair $(bc)$, $\ell_a$ that of the spectator,
		and the pair channel label $n_a=\{J_a,S_a,L_a\}$.  Applying the
		graphical recoupling rules of El-Baz and
		Castel~\cite{ElBazCastel1972}, the partial-wave projection of the
		kernel $Z^{ab}$ between the channel states~(\ref{eq:couplingB}) of
		partitions $a$ and $b$ reads
		\begin{align}
			Z^{J^{\pi}I}_{a n_a i,\;b n_b j}(q_a,q_b;E)
			&=(-1)^{R}(4\pi)^{-2}
			\bigl[\,\ell_a L_a S_a J_a\Sigma_a\,
			\ell_b L_b S_b J_b\Sigma_b\,
			I_a I_b J^{2}\bigr]
			\nonumber\\
			&\quad\times
			\begin{Bmatrix}
				t_a & I & I_a\\
				t_b & t_c & I_b
			\end{Bmatrix}
			\nonumber\\
			&\quad\times\sum_{L}[L^{2}](-1)^{L}\,
			F_{a n_a i,\,b n_b j;L}(q_a,q_b;E)
			\sum_{\mathcal{L}_a=0}^{L_a}\sum_{\mathcal{L}_b=0}^{L_b}
			\bigl[\,L_a\!-\!\mathcal{L}_a\;\;L_b\!-\!\mathcal{L}_b\,\bigr]
			\nonumber\\
			&\quad\times
			q_a^{\mathcal{L}_a+\mathcal{L}_b}\,
			q_b^{(L_a+L_b)-(\mathcal{L}_a+\mathcal{L}_b)}
			\left(\frac{\mu_a}{m_c}\right)^{\mathcal{L}_a}
			\left(\frac{\mu_b}{m_c}\right)^{L_b-\mathcal{L}_b}
			\nonumber\\
			&\quad\times
			\sqrt{\frac{(2L_a+1)!}{(2\mathcal{L}_a)!\,\bigl(2(L_a-\mathcal{L}_a)+1\bigr)!}}\;
			\sqrt{\frac{(2L_b+1)!}{(2\mathcal{L}_b)!\,\bigl(2(L_b-\mathcal{L}_b)+1\bigr)!}}
			\nonumber\\
			&\quad\times\sum_{f f_1 f_2}\bigl[f^{2}f_1^{2}f_2^{2}\bigr](-1)^{f}
			\begin{pmatrix}L_a\!-\!\mathcal{L}_a & L_b\!-\!\mathcal{L}_b & f_2\\ 0&0&0\end{pmatrix}
			\begin{pmatrix}\mathcal{L}_a & \mathcal{L}_b & f_1\\ 0&0&0\end{pmatrix}
			\nonumber\\
			&\quad\times
			\begin{pmatrix}f_1 & \ell_a & L\\ 0&0&0\end{pmatrix}
			\begin{pmatrix}f_2 & \ell_b & L\\ 0&0&0\end{pmatrix}
			\nonumber\\
			&\quad\times
			\begin{Bmatrix}f_2&f_1&f\\ \ell_a&\ell_b&L\end{Bmatrix}
			\begin{Bmatrix}\Sigma_b&f&\Sigma_a\\ \ell_a&J&\ell_b\end{Bmatrix}
			\begin{Bmatrix}
				L_a & L_a\!-\!\mathcal{L}_a & \mathcal{L}_a\\
				L_b & L_b\!-\!\mathcal{L}_b & \mathcal{L}_b\\
				f & f_2 & f_1
			\end{Bmatrix}
			\nonumber\\
			&\quad\times
			\begin{Bmatrix}
				\Sigma_a & s_a & S_b & L_b &\\
				J_a & s_c & J_b & f & 1\\
				L_a & S_a & s_b & \Sigma_b &
			\end{Bmatrix},
			\label{eq:VgeomB}
		\end{align}
		with the overall phase
		\begin{equation}
			R=(2s_c+s_b+S_b+L_a+J)+(2t_a+t_b+t_c+I_a),
			\label{eq:RphaseB}
		\end{equation}
		the bracket convention $[j_1j_2\ldots]=\sqrt{(2j_1+1)(2j_2+1)\cdots}$,
		the last symbol in Eq.~(\ref{eq:VgeomB}) being a $12j$ symbol of the
		first kind in the El-Baz--Castel classification, and the radial
		integral
		\begin{align}
			F_{a n_a i,\,b n_b j;L}
			&=\frac12\int_{-1}^{+1}\!\!d\xi\,P_L(\xi)\,
			\frac{p_1^{-L_a} g^a_i(p_1)\,
				g^b_j(p_2)\,p_2^{-L_b}}
			{E-\dfrac{q_a^{2}}{2\mu_b}-\dfrac{q_b^{2}}{2\mu_a}
				-\dfrac{q_a q_b\,\xi}{m_c}},
			\label{eq:FintB}
		\end{align}
		where
		\begin{align}
			p_1&=\sqrt{\frac{\mu_a^{2}}{m_c^{2}}q_a^{2}+q_b^{2}
				+\frac{2\mu_a}{m_c}q_a q_b\xi},
			\nonumber\\
			p_2&=\sqrt{q_a^{2}+\frac{\mu_b^{2}}{m_c^{2}}q_b^{2}
				+\frac{2\mu_b}{m_c}q_a q_b\xi}.
			\label{eq:p1p2B}
		\end{align}

		In Eq.~(\ref{eq:VgeomB}) the phase
		$(-1)^R$ contains the exchange sign $\zeta_{ab}$ through $L_a$
		and $J$.  The isospin $6j$ symbol is replaced by unity in a charge
		basis.  The Legendre sum over $L$ with the integral~(\ref{eq:FintB})
		is the only energy-dependent factor.  The double sum over
		$(\mathcal{L}_a,\mathcal{L}_b)$ is the solid-harmonic splitting of each pair
		orbital momentum between the two Jacobi momenta: the mass ratios raised to the powers $\mathcal{L}$ are the Jacobi coefficients of the
		momentum transformation, and the factorial square roots are the
		normalization of that expansion.  The four $3j$ symbols with zero
		projections enforce parity and triangle selection and annihilate most $(\mathcal{L},f)$ combinations, so that only a few terms survive in the multiple sum; the
		two $6j$ symbols recouple $(f_1f_2f)$ with $(\ell_a \ell_b L)$ and
		$(\Sigma_a\Sigma_b f)$ with $(\ell_a \ell_b J)$; the $9j$ symbol recouples
		the split orbital momenta; and the $12j$ symbol carries the spin chain
		$\Sigma,s,S,L,J,f$ across both partitions.  The factors $p^{-L_a}$,
		$p^{-L_b}$ in Eq.~(\ref{eq:FintB}) show that the kernel involves the
		reduced form factors $\hat g=g/p^{L}$: the threshold powers are
		generated by the $q^{\mathcal{L}}$ factors and mass ratios of
		Eq.~(\ref{eq:VgeomB}) and are not part of $\hat g$ itself.

		Discretized on a Gauss--Legendre mesh, the bound state solves
		$\det[\mathbb{1}-\mathcal{K}(E)]=0$ with
		$\mathcal{K}=W_{ab}\zeta_{ab}\,\tau^a Z^{ab} q_b^2 w_b$,
		equivalently the largest kernel eigenvalue $\eta_1(E)$ crossing
		unity.

	\section{$NN$ and $\alpha N$ potentials and their EST representations}
	\label{c}

	The nucleon--nucleon interaction is the Malfliet--Tjon (MT)
	potential~\cite{Malfliet1969} (two Yukawa terms) in its $^{1}S_{0}$
	(MT~I) and $^{3}S_{1}$ (MT~III) channels, and the $\alpha N$
	interaction is the Sack--Biedenharn--Breit (SBB)
	potential~\cite{Sack1954} (a Gaussian with a spin--orbit term in the
	$S$ and $P$ waves and a purely central Gaussian in the $D$ wave).  For
	the functional forms and the parameter values the reader is referred to
	the original papers; since there is documented confusion in the
	literature over the parameters of these simple model potentials, this
work uses
	the corrected values exactly as compiled in Table~I of
	Ref.~\cite{Schellingerhout1993}.

	The separable representations are constructed by the EST
	method~\cite{EST1973}: for a channel with support states
	$|\psi_i\rangle$ at energies $E_i$ (scattering states, plus the bound
	state where one exists),
	\begin{align}
		g_i &= V|\psi_i\rangle,
		\label{ff} \\
		\left[\bm{\Lambda}^{-1}\right]_{ij} &= \langle\psi_i|V|\psi_j\rangle,
		\label{power}
	\end{align}
	so that the representation is exact half-on-shell at every support
	energy.  The rank-4 support energies are $E_i=\{E_B,3,20,60\}$~MeV for
	the SBB $^{2}S_{1/2}$ channel ($E_B$ its bound state),
	$\{1,5,20,60\}$~MeV for the SBB $P$ and $D$ waves, $\{0.5,5,25,80\}$~MeV for the MT $^{1}S_{0}$, and $\{E_{d},1,10,50\}$~MeV
	for the MT $^{3}S_{1}$ ($E_{d}$ its deuteron bound state, which is
	physical and is not projected).

	For use in the three-body equations the exact half-shell form factors
	are fitted to the analytic form
	\begin{equation}
		g_i(p) = \frac{p^{L}}{(p^{2}+\beta^{2})^{m}}
		\sum_{n=0}^{10} a_{in}\,P_n(x),
		\qquad
		x=\frac{\beta^{2}-p^{2}}{\beta^{2}+p^{2}},
		\label{ffana}
	\end{equation}
	with $P_n$ the Legendre polynomials.  The variable $x$ maps the momentum half-axis onto the interval $[-1,1]$, with $x=0$ at $p=\beta$, so that the Legendre polynomials form a complete and numerically stable basis.
	The integer $m$ determines the power-law decrease of the ansatz at large momenta, $g_i(p)\sim p^{L-2m}$: it
	must be large enough for the loop integrals
	$\Delta_{ij}(E)=\langle g_i|G_0(E)|g_j\rangle$ to converge, and is
	chosen to reproduce the decay of the exact half-shell form factor.  For
	the potentials used here $m=1$ suffices in every channel, with
	$\beta=1.5$~fm$^{-1}$ (SBB $S$ wave), $1.2$~fm$^{-1}$ (SBB $P$ and $D$
	waves), and $2.0$~fm$^{-1}$ (MT $^{1}S_{0}$ and $^{3}S_{1}$).  The coupling matrix is rebuilt
	from the fitted factors (and symmetrized), so that the pair
	$(g,\bm\Lambda^{-1})$ is internally consistent at the level of the fit
	residual.  The Pauli-forbidden state of the $^{2}S_{1/2}$ channel is the
	bound state of the fitted representation itself, expressed in the
	two-pole form
	\begin{equation}
		\phi_f(p) =
		\frac{1}{(p^{2}+\beta^{2})(p^{2}+\gamma^{2})}
		\sum_{n=0}^{10} A_n P_n(x),
		\label{fbwf2}
	\end{equation}
	with $\gamma=0.669354$~fm$^{-1}$
	($\varepsilon_f=-11.611$~MeV, coinciding with the parent eigenvalue to
	four digits); using the bound state of the representation in use realizes
	the exactness condition of Sec.~3.  The complete coefficient sets $\{a_{in}\}$, $[\bm\Lambda^{-1}]_{ij}$, and
	$\{A_n\}$, for the rank-4 and rank-6 representations alike, are provided
	in tabulated and machine-readable form as ancillary files with the arXiv
	version of this article.

	The quality of the representations is summarized in
	Table~\ref{tab:quality}, which compares low-energy observables computed
	from the fitted EST tables with those of the parent potentials and with
	experiment.  In addition, the maximal deviation of the fitted phase
	shifts from those of the parent potentials over the interval
	$0.5$--$60$~MeV is $0.4^\circ$ for the MT $^{1}S_{0}$, $0.1^\circ$ for the MT $^{3}S_{1}$, $1.8^\circ$ for
	the SBB $^{2}S_{1/2}$ and $^{2}P_{3/2}$, $2.6^\circ$ for the SBB
	$^{2}P_{1/2}$, and $2.0^\circ$ for the SBB $D$ wave.  The representation error is one to two orders of
	magnitude smaller than the deviation of the models themselves from
	experiment, so the finite-rank expansion does not noticeably distort the
	two-body input.

	\begin{table}[H]
		\caption{Low-energy two-body observables of the fitted rank-4 EST
		representations, compared with the parent potentials and with
		experiment: scattering length $a$, effective range $r$, and the
		energy of the $S$-wave bound state (the $\alpha N$ forbidden state $\varepsilon_f$ and the MT deuteron $E_d$).  Experimental values from
		standard compilations; see, e.g., Ref.~\cite{Hlophe2017} and
		references therein.}
		\centering
		\begin{tabular}{@{}lccc@{}}
			\toprule
			 & EST & parent & experiment \\
			\midrule
			\multicolumn{4}{@{}l}{MT $^1S_0$} \\
			\quad $a$ (fm) & $-24.12$ & $-23.64$ & $-23.74$ \\
			\quad $r$ (fm) & $2.886$ & $2.892$ & $2.77$ \\
			\midrule
			\multicolumn{4}{@{}l}{MT $^3S_1$} \\
			\quad $E_{d}$ (MeV) & $-2.2324$ & $-2.2323$ & $-2.2246$ \\
			\quad $a$ (fm) & $5.497$ & $5.511$ & $5.42$ \\
			\quad $r$ (fm) & $1.894$ & $1.897$ & $1.76$ \\
			\midrule
			\multicolumn{4}{@{}l}{SBB $^{2}S_{1/2}$} \\
			\quad $\varepsilon_f$ (MeV) & $-11.6108$ & $-11.6105$ & --- \\
			\quad $a$ (fm) & $2.482$ & $2.483$ & $2.46$ \\
			\quad $r$ (fm) & $0.993$ & $0.990$ & --- \\
			\bottomrule
		\end{tabular}
		\label{tab:quality}
	\end{table}

		\bibliographystyle{unsrt} 
		\bibliography{references}
		
\end{document}